\documentstyle[12pt]{article}
\input amssym.def
\begin{document}
\begin{center}{\LARGE\bf{A Class of Bound Entangled States}}
\end{center}
\vspace{2ex}
\begin{center}
Shao-Ming Fei$^{1,2,3}$,\, Xianqing Li-Jost$^3$ and Bao-Zhi Sun$^1$
\bigskip

\begin{minipage}{5.2in}
$^1$ Department of Mathematics, Capital Normal University, Beijing 100037\\
$^2$ Institut f{\"u}r Angewandte Mathematik, Universit{\"a}t Bonn, 53115, Bonn\\
$^3$ Max-Planck-Institute for Mathematics in the Sciences, 04103 Leipzig\\
\end{minipage}
\end{center}

\parindent=18pt
\parskip=6pt
\begin{center}
\begin{minipage}{5in}
\vspace{3ex} \centerline{\large Abstract} \vspace{2ex}

We construct a set of PPT (positive partial transpose) states and
show that these PPT states are not separable,
thus present a class of bound entangled quantum states.

\end{minipage}
\end{center}

\bigskip
Keywords: Bound Entangled states, PPT

PACS: 03.65.Bz; 89.70.+c

\bigskip
\bigskip

Quantum entangled states are used as key resources in quantum information processing
such as quantum teleportation, cryptography, dense coding, error correction
and parallel computation \cite{2,3}. To quantify the degree of entanglement a
number of entanglement measures have been proposed for bipartite states.
However most proposed measures of entanglement
involve extremizations which are difficult to handle analytically.
It turns out that to verify the separability of a general
mixed states could be extremely difficult.

Among the quantum entangled states, there is a special kind of
states that can not be distilled. These states are called bound entangled states.
Many powerful separability criteria could not detect the entanglement
of these states, e.g. the bound entangled states given in \cite{UPB,HorodeckiPRL99,9}.
A few new bound entangled states have been found recently by using the method of positive
maps \cite{piani}.

It has been shown \cite{8} that any state which is entangled and satisfies
positive partial transpose (PPT) condition \cite{6} is not distillable.
The existence of PPT entangled states was discussed in \cite{7}
and explicit examples were provided in \cite{9}, based on
an elegant separability (range) criterion.

In \cite{10} a special class of quantum states ($d$-computable states)
were constructed. The entanglement of formation of these states can be analytically
calculated and it turns out that all the states are entangled.
In this paper according to the construction of the $d$-computable states,
we present first a class of PPT states, then by using
the range criterion we prove that they are entangled.

Let ${\cal{H}}$ be an $N-$dimensional complex Hilbert space with orthonormal basis
$e_i, i=1,\cdots,N$.
A general bipartite pure state on ${\cal{H}}\otimes{\cal{H}}$ is of the form,
\begin{equation}\label{psi}
|\psi>=\sum_{i,j=1}^Na_{ij}e_i\otimes e_j,\ \ \ \ \ \ \ \ \  a_{ij}\in{\cal{C}}
\end{equation}
with normalization $\sum\limits_{i,j=1}^Na_{ij}a_{ij}^*=1$. Let $A$ denote the
matrix with entries given by $a_{ij}$ in (\ref{psi}). Set
\begin{equation}\label{A}
A=\left [\matrix {
0 & b_1 & a & -c\cr
-b_1 & 0 & c & d\cr
-a & -c & 0 & -c_1\cr
c & -d & c_1 & 0}
\right ],
\end{equation}
where $a,c,d,b_1,c_1\in{\cal{C}}$.

It was shown that all pure states $|\psi>$ with $A$ given by (\ref{A}) have
a simple formula of the generalized concurrence $C$, such that the entanglement
of formation is a monotonically increasing function of $C$ \cite{10}. Moreover the
entanglement of formation of all mixed states $\rho$ with decompositions on pure
states with $A$ given by (\ref{A}) can be analytically calculated. As all the states
with $A$ of (\ref{A}) are entangled, these mixed states $\rho$ are also entangled.

In fact $A$ is an antisymmetric matrix. Any antisymmetric matrices
are equivalent to the following standard form under similarity
transformations:

~~~~~~~~~~~~~~~~~~~~~$A_1= \left[\matrix{ 0&\lambda_1&0&0\cr -\lambda_1&0&0&0\cr
0&0&0&\lambda_2\cr 0&0&-\lambda_2&0} \right]$~~ or ~~
$A_2=
\left[\matrix{
0&0&\lambda_1&0\cr
0&0&0&\lambda_2\cr
-\lambda_1&0&0&0\cr
0&-\lambda_2&0&0}
\right].
$

If we set $\lambda_1=\pm b$, $\lambda_2=-c$ in $A_1$, the matrix $A_1$ gives rise to
two pure states
$$|\psi_{\pm b}>=\left[\matrix{
0&\pm {b}&0&0&\mp {b}&0&0&0&0&0&0&-{c}&0&0&{c}&0}
\right]^t,$$ and hence two projectors
$\rho _{\pm b}=|\psi_{\pm b}><\psi_{\pm b}|$,
where $t$ denotes the transposition.
We define
$$\rho_b=\frac{1}{2}\rho_{+b}+\frac{1}{2}\rho_{-b}.$$

And if we set $\lambda_1=\pm a$, $\lambda_2=d$ in $A_2$ we have
$$\rho_a=\frac{1}{2}\rho_{+a}+\frac{1}{2}\rho_{-a},$$
where $\rho_{\pm a}=|\psi_{\pm a}><\psi_{\pm a}|$,
$$|\psi_{\pm a}>=\left[\matrix{
0&0&\pm {a}&0&0&0&0& {d}&\mp {a}&0&0&0&0& -{d}&0&0}
\right]^t.$$
We define
\begin{equation}
\rho_0=\frac{1}{2}\rho_{a}+\frac{1}{2}\rho_{b}
\end{equation}
The state $\rho_0$ is not separable as its partial transposed matrix has a
negative eigenvalue. Below we mix
$\rho_0$ with some separable states in such a way that the resulting
state will be both partial transposition positive and entangled.
Let $I_4$ be a $16\times 16$ matrix with only non-zero entries
$(I_4)_{1,1}=(I_4)_{6,6}=(I_4)_{11,11}=(I_4)_{16,16}=1$. We consider
$\rho=(1-\varepsilon)I_4+\varepsilon\rho_0$,
which is of the form:
\begin{equation}\label{rho}
\rho={\small{\left[\matrix{
x_1&0&0&0&0&0&0&0&0&0&0&0&0&0&0&0\cr
0&x_3&0&0&-x_3&0&0&0&0&0&0&0&0&0&0&0\cr
0&0&x_2&0&0&0&0&0&-x_2&0&0&0&0&0&0&0\cr
0&0&0&0&0&0&0&0&0&0&0&0&0&0&0&0\cr
0&-x_3&0&0&x_3&0&0&0&0&0&0&0&0&0&0&0\cr
0&0&0&0&0&x_1&0&0&0&0&0&0&0&0&0&0\cr
0&0&0&0&0&0&0&0&0&0&0&0&0&0&0&0\cr
0&0&0&0&0&0&0&x_5&0&0&0&0&0&-x_5&0&0\cr
0&0&-x_2&0&0&0&0&0&x_2&0&0&0&0&0&0&0\cr
0&0&0&0&0&0&0&0&0&0&0&0&0&0&0&0\cr
0&0&0&0&0&0&0&0&0&0&x_1&0&0&0&0&0\cr
0&0&0&0&0&0&0&0&0&0&0&x_4&0&0&-x_4&0\cr
0&0&0&0&0&0&0&0&0&0&0&0&0&0&0&0\cr
0&0&0&0&0&0&0&-x_5&0&0&0&0&0&x_5&0&0\cr
0&0&0&0&0&0&0&0&0&0&0&-x_4&0&0&x_4&0\cr
0&0&0&0&0&0&0&0&0&0&0&0&0&0&0&x_1}
\right]}},
\end{equation}
where $x_1=\frac{1-\varepsilon}{4}$, $x_2=\frac{\varepsilon}{2}|a|^2$,
$x_3=\frac{\varepsilon}{2}|b|^2$,
$x_4=\frac{\varepsilon}{2}|c|^2$,
$x_5=\frac{\varepsilon}{2}|d|^2$.

The eigenvalues of the partial transposed matrix with respect to the
second subspace $\rho^{T_2}$ of $\rho$ are
$$
0,0,0,0,\frac{\varepsilon}{2}|a|^2,\frac{\varepsilon}{2}|a|^2,
\frac{\varepsilon}{2}|b|^2,\frac{\varepsilon}{2}|b|^2,
\frac{\varepsilon}{2}|c|^2,\frac{\varepsilon}{2}|c|^2,
\frac{\varepsilon}{2}|d|^2,\frac{\varepsilon}{2}|d|^2
$$
together with the roots of the following equation:
\begin{equation}\label{4}
(\lambda-\frac{1-\varepsilon}{2})^4
-\frac{\varepsilon^2}{4}(|a|^4+|b|^4+|c|^4+|d|^4)(\lambda-\frac{1-\varepsilon}{4})^2
+\frac{\varepsilon^4}{16}(|a|^2|d|^2-|b|^2|c|^2)^2=0,
\end{equation}
i.e.
$\lambda=\frac{1-\varepsilon}{4}\pm\frac{\varepsilon}{4}
\sqrt{2[(|a|^4+|b|^4+|c|^4+|d|^4)\pm\sqrt{\Delta_1}]}$,
where $\Delta_1$ is defined by the discriminant of (\ref{4}):
$$
\Delta=\frac{\varepsilon^4}{16}
[(|a|^2-|d|^2)^2+(|b|^2+|c|^2)^2]
[(|a|^2+|d|^2)^2+(|b|^2-|c|^2)^2]
=\frac{\varepsilon^4}{16}\Delta_1.
$$
If
\begin{equation}\label{cond}
\frac{1-\varepsilon}{4}-\frac{\varepsilon}{4}
\sqrt{2[(|a|^4+|b|^4+|c|^4+|d|^4)+\sqrt{\Delta_1}]}>0,
\end{equation}
then $\rho^{T_2}$ is positive semidefinite.
It is possible that the condition (\ref{cond}) is satisfied while
keeping the state $\rho_4$ entangled. For simplicity we
take $a=b=c=d=\frac{1}{2}$. In this case $\rho^{T_2}$ is positive semidefinite
when $0\leq\varepsilon\leq\frac{1}{2}$, and the state becomes
\begin{equation}\label{rho4}
\rho={\scriptsize{\left[\matrix{
\displaystyle\frac{1-\varepsilon}{4}&0&0&0&0&0&0&0&0&0&0&0&0&0&0&0\cr
0&\displaystyle\frac{\varepsilon}{8}&0&0&-\displaystyle\frac{\varepsilon}{8}&0&0&0&0&0&0&0&0&0&0&0\cr
0&0&\displaystyle\frac{\varepsilon}{8}&0&0&0&0&0&-\displaystyle\frac{\varepsilon}{8}&0&0&0&0&0&0&0\cr
0&0&0&0&0&0&0&0&0&0&0&0&0&0&0&0\cr
0&-\displaystyle\frac{\varepsilon}{8}&0&0&\displaystyle\frac{\varepsilon}{8}&0&0&0&0&0&0&0&0&0&0&0\cr
0&0&0&0&0&\displaystyle\frac{1-\varepsilon}{4}&0&0&0&0&0&0&0&0&0&0\cr
0&0&0&0&0&0&0&0&0&0&0&0&0&0&0&0\cr
0&0&0&0&0&0&0&\displaystyle\frac{\varepsilon}{8}&0&0&0&0&0&-\displaystyle\frac{\varepsilon}{8}&0&0\cr
0&0&-\displaystyle\frac{\varepsilon}{8}&0&0&0&0&0&\displaystyle\frac{\varepsilon}{8}&0&0&0&0&0&0&0\cr
0&0&0&0&0&0&0&0&0&0&0&0&0&0&0&0\cr
0&0&0&0&0&0&0&0&0&0&\displaystyle\frac{1-\varepsilon}{4}&0&0&0&0&0\cr
0&0&0&0&0&0&0&0&0&0&0&\displaystyle\frac{\varepsilon}{8}&0&0&-\displaystyle\frac{\varepsilon}{8}&0\cr
0&0&0&0&0&0&0&0&0&0&0&0&0&0&0&0\cr
0&0&0&0&0&0&0&-\displaystyle\frac{\varepsilon}{8}&0&0&0&0&0&\displaystyle\frac{\varepsilon}{8}&0&0\cr
0&0&0&0&0&0&0&0&0&0&0&-\displaystyle\frac{\varepsilon}{8}&0&0&\displaystyle\frac{\varepsilon}{8}&0\cr
0&0&0&0&0&0&0&0&0&0&0&0&0&0&0&\displaystyle\frac{1-\varepsilon}{4}\cr
}\right]}}.
\end{equation}

Recall that if a state $\rho$ acting on Hilbert space
${\cal{H}=\cal{H}}\otimes{\cal{H}}$ is separable, then there
exists a set of product vectors $\{|\psi_i>\otimes|\phi_k>\}$,
$\{i,k\}\in I$ ($I$ is a finite set of pairs of indices with
number of pairs $M=\#I\leq N^2$) and probabilities $p_{ik}$ such
that the ensemble $\{|\psi_i>\otimes|\phi_k>,p_{ik}\}$
($\{|\psi_i>\otimes|\phi_k^*>,p_{ik}\}$) corresponds to the matrix
$\rho$ ($\rho^{T_2}$), and the vectors
$\{|\psi_i>\otimes|\phi_k>\}$ ($\{|\psi_i>\otimes|\phi_k^*>\}$)
span the range of $\rho$ ($\rho^{T_2}$). In particular any vector
$\{|\psi_i>\otimes|\phi_k>\}$ ($\{|\psi_i>\otimes|\phi_k^*>\}$)
belongs to the range of $\rho$ ($\rho^{T_2}$), see \cite{9}.

Now we calculate all the product (unnormalised) vectors belonging
to the range of $\rho$. With the basis ordered in the following way
$e_1\otimes e_1,\ e_1\otimes e_2,\ e_1\otimes e_3,\  e_1\otimes e_4,
\ e_2\otimes e_1,\ e_2\otimes e_2,\ \cdots, \ e_4\otimes e_4$,
any vector belonging to the range of $\rho$ can be presented as
\begin{equation}\label{5}
\mu=\left[\matrix{
A&B&C&0&-B&D&0&E&-C&0&F&G&0&-E&-G&H
}\right]^t,
\end{equation}
where $A,B,C,D,E,F,H\in{\cal C}$.
On the other hand a separable $\mu$ is of the form
\begin{equation}\label{qq}
\mu_{sep}=
\left[\matrix{b_1&b_2&b_3&b_4}\right]^t
\otimes
\left[\matrix{c_1&c_2&c_3&c_4}\right]^t,
\end{equation}
$b_1,b_2,b_3,b_4,c_1,c_2,c_3,c_4\in{\cal C}$.
Comparing (\ref{qq}) with (\ref{5}) we have
\begin{eqnarray}
&&b_1c_4=b_2c_3=b_3c_2=b_4c_1=0\label{7},\\
&&b_1c_2=-b_2c_1\label{8},\\
&&b_1c_3=-b_3c_1\label{9},\\
&&b_2c_4=-b_4c_2\label{10},\\
&&b_3c_4=-b_4c_3\label{11}.
\end{eqnarray}
To find a set of basic separable vectors that span the range of (\ref{5}),
let us consider the following cases:

I) $b_1b_2\neq0$. Without loss of generality we set $b_1=1$
and $c_1=A,\ c_2=B$. From (\ref{7}), we have $c_4=c_3=0$ and $b_3=b_4=0$.
From (\ref{8}) we have $b_2=-\frac{B}{A}$, and
$b_2c_2=D$ by comparing (\ref{5}) with (\ref{qq}).
Therefore $B^2=-AD$, and $B=\pm\sqrt{-AD}$.
Hence we have $b_2=\mp\frac{\sqrt{-AD}}{A},\ c_2=\pm\sqrt{-AD}$.
Thus we obtain the states
\begin{equation}\label{12}
\frac{1}{A}\left[\matrix{A&-\sqrt{-AD}&0&0}\right]^t
\otimes
\left[\matrix{A&\sqrt{-AD}&0&0}\right]^t
\end{equation}
and
\begin{equation}\label{13}
\frac{1}{A}\left[\matrix{A&\sqrt{-AD}&0&0}\right]^t
\otimes
\left[\matrix{A&-\sqrt{-AD}&0&0}\right]^t.
\end{equation}

II) $b_1b_2=0$

\ \ \ i) $b_1\neq0,\ b_2=0$. We set $b_1=1$.

\ \ \ \ \ \ If $b_3b_4\neq0$ or $b_3=0,\ b_4\neq0$, then only the null vector
satisfies these conditions.

\ \ \ \ \ If $b_3=b_4=0$, from (\ref{5}) and (\ref{qq}), we obtain
$c_1=A,\ c_2=c_3=c_4=0$ and the vector is of the form
\begin{equation}\label{14}
A\left[\matrix{1&0&0&0}\right]^t\otimes\left[\matrix{1&0&0&0}\right]^t.
\end{equation}

\ \ \ \ \ \ If $b_3\neq0,\ b_4=0$, then similar to the case I),
we have the following vectors :
\begin{equation}\label{15}
\frac{1}{A}\left[\matrix{A&0&-\sqrt{-AF}&0}\right]^t
\otimes
\left[\matrix{A&0&\sqrt{-AF}&0}\right]^t
\end{equation}
\ \ \ \ \ \ \ \ \ \ \ \ \ \ \ \ and
\begin{equation}\label{16}
\frac{1}{A}\left[\matrix{A&0&\sqrt{-AF}&0}\right]^t
\otimes
\left[\matrix{A&0&-\sqrt{-AF}&0}\right]^t.
\end{equation}

\ \ \ ii) $b_1=0,\ b_2\neq0$. We take $b_2=1$. Similar to the
previous case, we have the following two cases:

\ \ \ \ \ \ If $b_3=b_4=0$, then $c_1=c_3=c_4=0,\ c_2=D$,
the vector is
\begin{equation}\label{19}
D\left[\matrix{0&1&0&0}\right]^t
\otimes
\left[\matrix{0&1&0&0}\right]^t.
\end{equation}

\ \ \ \ \ \ If $b_3=0,\ b_4\neq0$ then we have
\begin{equation}\label{22}
\frac{1}{D}\left[\matrix{0&D&0&-\sqrt{-DH}}\right]^t
\otimes
\left[\matrix{0&D&0&\sqrt{-DH}}\right]^t
\end{equation}
\ \ \ \ \ \ \ \ \ \ \ \ \ \ \ \ and
\begin{equation}\label{23}
\frac{1}{D}\left[\matrix{0&D&0&\sqrt{-DH}}\right]^t
\otimes
\left[\matrix{0&D&0&-\sqrt{-DH}}\right]^t.
\end{equation}

\ \ \ iii) $b_1=b_2=0$

\ \ \ \ \ \ If $b_3b_4\neq0$, taking $b_3=1$, we have
\begin{equation}\label{24}
\frac{1}{F}\left[\matrix{0&0&F&-\sqrt{-FH}}\right]^t
\otimes
\left[\matrix{0&0&F&\sqrt{-FH}}\right]^t
\end{equation}
\ \ \ \ \ \ \ \ \ \ \ \ \ \ \ \ and
\begin{equation}\label{25}
\frac{1}{F}\left[\matrix{0&0&F&\sqrt{-FH}}\right]^t
\otimes
\left[\matrix{0&0&F&-\sqrt{-FH}}\right]^t.
\end{equation}

\ \ \ \ \ \ If $b_3\neq0,\ b_4=0$, taking $b_3=1$, we obtain
\begin{equation}\label{26}
F\left[\matrix{0&0&1&0}\right]^t
\otimes
\left[\matrix{0&0&1&0}\right]^t.
\end{equation}

\ \ \ \ \ \ If $b_3=0,\ b_4\neq0$, taking $b_4=1$, we get
\begin{equation}\label{27}
H\left[\matrix{0&0&0&1}\right]^t
\otimes
\left[\matrix{0&0&0&1}\right]^t.
\end{equation}

The vectors (\ref{12}), (\ref{13}), (\ref{14}) and (\ref{19}) are
linear dependent. So we can exclude (\ref{13}). For the same
reason, we can remove (\ref{16}),  (\ref{23}) and (\ref{25}). The
left vectors (\ref{12}), (\ref{14}), (\ref{15}), (\ref{19}),
 (\ref{22}), (\ref{24}), (\ref{26}) and (\ref{27}) span
the separable linear independent vectors of the range of $\rho$.
The partial complex conjugations (PCC) of these vectors, e.g. PCC
of (\ref{12}),
$\frac{1}{A}\left[\matrix{A&-\sqrt{-AD}&0&0}\right]^t
  \otimes\left[\matrix{A^*&\sqrt{-AD}^*&0&0}\right]^t$,
do not span the range of $\rho^{T_2}$ as the vector
\begin{equation}
\left[\matrix{1&0&0&0}\right]^t\otimes\left[\matrix{0&1&0&0}\right]^t,
\end{equation}
which does belong to the range of $\rho^{T_2}$, does not instead belong to their linear span.
Hence for any $0\leq k\leq\frac{1}{2}$ the state $\rho$
violates the separability criterion in \cite{9}. Thus the states (\ref{rho4}) are bound entangled ones.

We have provided a class of inseparable states with positive partial transposition
by using the range criterion. Although we have taken $a=b=c=d=\frac{1}{2}$
for simplicity, in fact, the state (\ref{rho}) is bound entangled as long as $\varepsilon$ is small enough such
that all the roots of (\ref{4}) are positive.

It is verified that the trace norm of the realigned matrix of (\ref{rho4}) is one. Hence
the realignment separability criterion \cite{Rudolph02} could not detect the entanglement of this
bound entangled state. Moreover the trace norm of $\rho^{T_2}$ is also one. Therefore neither
the lower bound of concurrence nor the lower bound for the entanglement of formation
\cite{11} could detect the entanglement.

\vspace{0.5truecm}
\noindent {\bf Acknowledgments}\,  We thank the referee for pointing out
a mistake in the first version.
The work is partially supported by NKBRPC (2004CB318000).


\begin{thebibliography}{99}
\bibitem{2}
M.A. Nielsen and I.L. Chuang, Quantum Computation and Quantum Information (Cambridge
University Press, Cambridge, 2000.

\bibitem{3}
D. Bouwmeester, A. Ekert and A. Zeilinger(Eds.), The Physics of Quantum Information:
Quantum Cryptography, Quantum Teleportation and Quantum Computation (Springer, New
York, 2000).

\bibitem{UPB} C.H. Bennett \textit{et al.}, Phys. Rev. Lett. \textbf{82},
5385 (1999).

\bibitem{HorodeckiPRL99} P. Horodecki, M. Horodecki, and R. Horodecki, Phys.
Rev. Lett. \textbf{82}, 1056 (1999).

\bibitem{9}
P. Horodecki, Phys. Lett. A, 232, 233 (1997).

\bibitem{piani}
M. Piani, {\it A class of $2^N \times 2^N$ bound entangled states revealed by non-decomposable maps},
quant-ph/0411098.

\bibitem{8}
M. Horodecki, P. Horodecki and R. Horodecki, Phys.Rev. Lett. 80, 5239 (1998);\\
P. Horodecki, M. Horodecki and R. Horodecki, Phys. Rev. Lett. 82, 1056 (1999).

\bibitem{6}
A. Peres, Phys. Rev. Lett. 77, 1413 (1996).

\bibitem{7}
M. Horodecki, P. Horodecki, and R. Horodecki, Phys.Lett. A 223, 1 (1996).

\bibitem{10}
S.M. Fei and X.Q. Li-Jost, Rep. Math. Phys. 53, 195(2004);\\
S.M. Fei, J. Jost, X.Q. Li-Jost and G.F. Wang, Phys. Lett. A 310, 333-338(2003).

\bibitem{Rudolph02} O. Rudolph, quant-ph/0202121;\\
K. Chen and L.A. Wu, Quant. Inf. Comp. 3, 193(2003).

\bibitem{11}
K. Chen, S. Albeverio and S.M. Fei, Phys. Rev. Lett. 95, 040504(2005).\\
K. Chen, S. Albeverio and S.M. Fei, Phys. Rev. Lett. 95, 210501(2005).


\end{thebibliography}
\end{document}